# Stocks and Cryptocurrencies: Antifragile or Robust? A Novel Antifragility Measure of the Stock and Cryptocurrency Markets

Darío Alatorre,[1,2] Carlos Gershenson,[1,3] and José L. Mateos[1,4]

[1] Centro de Ciencias de la Complejidad, Universidad Nacional Autónoma de México, Ciudad de México, C. P. 04510, México.
[2] Instituto de Matemáticas, Universidad Nacional Autónoma de México, Ciudad de México, C. P. 04510, México.
[3] Departamento de Ciencias de la Computación, Instituto de Investigaciones en Matemáticas Aplicadas y en Sistemas, Universidad Nacional Autónoma de México, Ciudad de México, C. P. 04510, México.
[4] Instituto de Física, Universidad Nacional Autónoma de México, Ciudad de México, C. P. 04510, México.

## Abstract

In contrast with robust systems that resist noise or fragile systems that break with noise, antifragility is defined as a property of complex systems that benefit from noise or disorder. Here we define and test a simple measure of antifragility for complex dynamical systems. In this work we use our antifragility measure to analyze real data from return prices in the stock and cryptocurrency markets. Our definition of antifragility is the product of the return price and a perturbation. We explore different types of perturbations that typically arise from within the system. Our results suggest that for both the stock market and the cryptocurrency market, the tendency among the 'top performers' is to be robust rather than antifragile. It would be important to explore other possible definitions of antifragility to understand its role in financial markets and in complex dynamical systems in general.

## 1. Introduction

Some systems have the property of being *robust*, in the sense that they are not affected by the presence of noise. An analogous concept of robustness is stability, where a complex dynamical system recovers its original stable state after a small perturbation. On the other extreme, we have the property of being *fragile*, where the presence of noise breaks or collapses the system. The case of fragility corresponds to a state in unstable equilibrium, where a perturbation drives the system away from the original state.

Now, let us describe the counterintuitive concept of *antifragility,* which is different from robustness and fragility. A system is antifragile when it benefits from noise or disorder. Thus, noise acts as a constructive element that allows the system to perform a function successfully.

Therefore, one expects to encounter antifragility in a complex dynamical system that is nonlinear, has many degrees of freedom and is out of equilibrium.

The term antifragility was coined by Taleb in his book of the same name [1]. Taleb's ideas have been further explored in different contexts not only in risk analysis and financial systems [2-4], but also as a means for strategic design and planning [5-7]. The original formal definition [8] (see also [9,10]), however, is difficult to use in practice and with real data. Therefore, we propose in this paper a novel definition of antifragility that can be implemented easily for a quantitative analysis and can be quite general.

This definition was inspired by some previous recent work, where it was proposed a pragmatic antifragility measure for complex systems, provided one can define measures of "satisfaction" for each of its agents and "perturbation" for the whole system. These measures are not supposed to be dependent on each other, as in Taleb's original idea, which assumes that the satisfaction is a function that depends on the magnitude of the perturbation. In this work, antifragility is defined as the product of perturbation and satisfaction. Consequently, an agent is robust if its antifragility is close to 0 and fragile if it is negative. The agent is antifragile for positive values of the measure. This new definition of antifragility has been studied recently in the context of random and biological Boolean networks [11], multilayer random Boolean networks [12], convolutional neural networks [13], and ecosystems [14].

In this work we tested this antifragility measure with real data from the stock market and the cryptocurrency market. In this context of economics and finance, we used the price returns of assets as a measure of satisfaction, while a perturbation measure for each system was defined as the mean of different volatility parameters coming from within the markets –involving the price returns, the market capitalization and the volume (amount) of transactions of each asset, as well as a global mean index and the volatility index VIX (the stock symbol and the popular name for the Chicago Board Options Exchange's CBOE Volatility Index, a popular measure of the stock market's expectation of volatility based on S&P 500 index options). The antifragility measure was calculated as the price return multiplied by the perturbation. We evaluated this antifragility measure in daily, weekly, monthly and semiannual time-scales, and compared it with different "good-performance" measures that we could extract from the data.

The identification of antifragility could be helpful in a wide range of scenarios. Particularly in economic and financial systems, which Taleb [1] suggested that should be antifragile. This work aims to apply Taleb's ideas to real data of the stock and cryptocurrencies market.

The paper is organized as follows: The next section gives some background of other physical systems that "gain from disorder", and the role of randomness and noise in financial systems. Then, we explain our definition of antifragility in general and in the context of stock and cryptocurrency markets. In the following sections, we describe the data and the variables employed and the results. Finally, we end with a discussion and concluding remarks.

## 2. Background

In the introduction of this paper, we mentioned that antifragile systems gain in the presence of noise. This noise can come from an external perturbation in an open system or can be an internal noise due to the complexity of the system itself. However, this idea has been known in the field of complex dynamical systems, nonlinear stochastic dynamics and statistical physics for more than 40 years. Thus, the concept of antifragility can be traced back to the studies of the constructive role of noise in these nonlinear systems out of equilibrium. We will very briefly describe some of these phenomena that have been thoroughly studied in the literature: stochastic resonance, noise-induced transport and optimization by simulated annealing. All these phenomena share the property of gaining in the presence of a stochastic signal or perturbation, thermal noise or by being exposed to fluctuations.

### 2.1 Constructive role of noise in the physical sciences

*Stochastic resonance.-* The idea of stochastic resonance is to add noise to a signal and measure the signal to noise ratio (SNR). Contrary to what is observed in linear systems, where the addition of noise decreases the SNR, some nonlinear systems exhibit an increase of the SNR as the intensity of the noise is increased; the signal reaches  a maximum after which the SNR starts to decrease. So, a SNR is a function of the intensity of the noise and has a maximum, similar to a resonance in a physical system; thus, the name of stochastic resonance. It is worth mentioning that the phenomenon of stochastic resonance is manifested when the system is driven out of equilibrium by applying an external force. Thus, even though the original system can be in equilibrium without any external force or perturbation, once we perturb the system, the full system (the original one plus the external forcing) can be considered as an out-of-equilibrium system. This phenomenon can be observed in physical and biological systems alike. The idea was introduced in the early 1980s by G. Parisi and collaborators. For a thorough review see [15]. Therefore, adding noise is beneficial and, in this sense, is closely related with the idea of antifragility.

*Noise-induced transport and Brownian motors.-* These phenomena refer to the counterintuitive idea of incorporating noise or a stochastic signal to enhance and generate transport in an asymmetric physical system. The asymmetry  can be spatial or temporal and it rectifies a symmetric noise generating a finite current. The phenomenon was studied by Richard Feynman in the early 1960s and gained attention in the 1990s with the study of thermal ratchets and Brownian motors that use Brownian motion on top of asymmetric ratchet potentials. One of the key motivations was to understand the mechanism behind motor proteins inside the cell and applications to nanoscience. For a review, see [16]. For a combination of Brownian motors and stochastic resonance, see [17]. Again, the main idea is that adding noise can be beneficial, as in an antifragile system.

*Optimization by simulated annealing.-* This is an important phenomenon where one implements simulated annealing, as in statistical mechanics, using thermal noise to escape from local minima in a landscape to arrive at a global minimum; thereby finding an optimization function. There is a vast literature around this idea. See for instance the original paper [18] and also [19].

There are other phenomena related to the idea of antifragility, like noise-induced synchronization in non-linear physical and biological systems [20], the "order from noise" [21], and the "complexity from noise" [22] principles.

### 2.2 Noise and fluctuations in economics and finance

Other important examples of complex dynamical systems that might be antifragile are economic and financial systems. Clearly, both are complex systems that are open, nonlinear, and have many agents that operate collectively, giving rise to emergent properties like the ones we observe in financial markets and other macroeconomic indicators. The question is if we can find the phenomenon of antifragility in economics and finance. Recent developments have used different definitions of antifragility to study recovery [23], collaborative business ecosystems [24] and the purchasing and supply chain management [25], but all of them go back to Taleb's original definition. Our approach is more heuristic in the sense that it aims at the identification and measurement of antifragility.

#### 2.2.1 Noise

Let us start with noise. In the context of economics and financial markets, the idea of noise can have different meanings. Noise can emerge as a result of a large number of small events due to the collective actions of economic agents on a daily basis. On the other hand, a small number of extreme events, like crashes, can have severe consequences as well. Noise makes trading in financial markets possible, and thus allows us to observe prices for financial assets. Noise causes markets to be somewhat inefficient, but often prevents us from taking advantage of inefficiencies. Noise in the form of uncertainty about the actions to be taken by a large number of agents in the future makes it very difficult for us to adopt rational decisions, and act by guessing or intuition instead. As a result, we tend to consider price fluctuations and price return in financial markets as a random or stochastic process, and act mainly by guessing. We are forced to act largely in the dark. See the classical paper of 1986 by Fisher Black titled simply "Noise" [26]. The topic of noise has a vast literature and extends in many directions, from mathematics and natural sciences, all the way to medicine, psychology, economics, politics, and so on. See the very recent book by Daniel Kahneman titled "Noise: A flaw in human judgment" [27].

#### 2.2.2 Perturbations

Regarding the perturbations that we use in this paper, we consider only internal perturbations (or noise) that comes from inside the market itself. This endogenous noise of the markets include the fluctuations observed in the returns of prices, volatility and market capitalization. As mentioned before, this noise emerges from the collective activities of many economic agents. Many of these agents do not have information and sometimes are called "noise traders" after the classical description by Black [26]. See also the book by Sornette [28], where different models are analyzed. These noise traders, among other internal considerations due to the complexity of markets, generate smaller scale shocks. However, financial markets can be affected by aggregate shocks as well; see [29,30]. These aggregate shocks arise due to external or exogenous events to the market, such as wars, terrorist attacks, massive migrations, political turmoils, or a pandemic. In our study here, we are not considering the effect of these aggregate shocks, but it would be important to take them into account. That would be an interesting extension to consider for future research.

#### 2.2.3 Price returns

Finally, let us discuss the last piece needed to construct our antifragility measure. We will see that this measure is the product of price returns and perturbations.

Price returns are one of the central quantities to study in the financial markets. It refers to the difference in prices of an asset evaluated at two different times. It can be a time difference of seconds, days, weeks, years or decades. There are several definitions of these price returns that can be ratios or percentage of price fluctuations, or differences between the logarithm of price (the so-called log returns). This central quantity is considered to be a random variable, according to the discussion above, and is therefore natural to study its statistical properties. In particular, it is important to study the probability distribution of price returns. An important result that has been established is that the stock returns probability distribution has a power law or fat tail characterized by an exponent close to three, the so-called cubic law (see [31] for a recent review of this cubic law and other power laws). These power-law tails in the distribution have important consequences in the dynamics of the price returns: due to these long tails, we expect to have large fluctuations and extreme events in the time series. These extreme events correspond to crashes in the stock market and, from the viewpoint of the statistical physics of complex systems, can be understood as a scale-free dynamics near criticality [28]. This scale-free dynamics implies that the system is prone to have large fluctuations and, in this sense, becomes less stable. The consequences of the power-law distributions of the returns have triggered a whole new field of study called *econophysics* [32,33], where there is an emphasis in trying to understand the origin of this long-range fat-tails distributions. The main idea is to model economics and finance as a complex system using the tools of statistical physics and nonlinear dynamics: for instance, using agent-based models as a "microscopic" description to explain the emergent phenomena observed in financial markets. At the end, the scale-free power laws observed in the return distribution implies some sort of phase transition where the system can become unstable (or metastable) locally and starts to develop a long-range dynamics.

### 2.2.4 Complexity in Economics and Finance

It is worth mentioning briefly new developments of complexity science in economics and finance. There is a vast literature around the topic. Recent papers dealing with the collective dynamics of stock markets, universal features and complexity in financial data are [34-37] and, in particular for the analysis of cryptocurrency markets [38-43].

Finally, the application of the emergent science of networks to financial systems is an important field of research; for a very recent review see [44] and references therein. Other important reviews on the foundations of complexity economics are [45] and in economic complexity theory and applications [46]. See also the book by Hidalgo [47].

It would be interesting to consider for future research the incorporation of the concept of antifragility as a feature of complex dynamical systems in general, and in particular from the viewpoint of complexity and networks in economics and finance.

## 3. Materials and Methods

We now describe the data set that we used throughout this paper.
We analyzed a data set [51] with historical daily data (open price and volume) of about 7,000 stocks from the US Stock Market. Such a dataset considered only active stocks by the end of 2017 and there is no data from stocks that disappeared from the market before that date. We truncated the data starting from 1990.

In the case of cryptocurrencies, the data [52] consist of daily historical data (open price, market capitalization and volume) of around 1800 assets, the oldest one starts in 2013 and ends in November 2018.
The raw data may be found in [51,52]. And all the processed data is available in this purl: https://purl.prod.archive.org/purl/AFSCM_data/processed-data.

### 3.1 A measure of satisfaction

We will start our analysis with the variation in time of the price of a stock or cryptocurrency. Let $p_i$ denote the price of a stock, and $q_i$ indicate the price of a cryptocurrency, at time ***i***, where ***i=1,2,...,N*** is a discrete variable indicating the time when the price is measured. The stock return is the difference between the prices of two consecutive times. That is:

$$S(i) = p_i - p_{i-1} \tag{1}$$

This quantity, the stock return, can be considered as a measure of "satisfaction", being positive when the price increases and negative otherwise. We will refer sometimes to the stock return in Eq. (1) as "satisfaction" in order to connect with a set of previous papers of one of the authors (CG), in which the concept of satisfaction was used –together with this definition of antifragility– for different fields such as ecology and networks (see [11-13]).

### 3.2 Definition of antifragility

In order to quantify how sensible the stock return of a given asset is to an external perturbation, we introduce here a definition of antifragility. We consider markets with stock returns $S$ and an external time-dependent perturbation $P$ . Let ***S(x,i)*** be the stock return of asset ***x*** at time ***i***, and ***P(i)*** be the *perturbation* measure of the system at time ***i***. We define the antifragility of asset ***x*** at time ***i*** as:

$$A(x, i) = S(x, i) \cdot P(i). \tag{2}$$

The global *antifragility* of asset $x$ is then defined as the mean value of ***A(x,i)*** over the whole time interval under consideration, *i.e.*

$$A(x) = mean_i(A(x, i)), \tag{3}$$

where $\boldsymbol{mean_i}$ refers to the arithmetic mean over index ***i***

$$mean_i(a_i) = \frac{1}{n}(a_1 + ... + a_n). \tag{4}$$

This definition of the concept of antifragility allows us to classify the systems as antifragile, robust and fragile depending on the overall result of the stock returns and the perturbation. We consider that a stock is antifragile if the stock return is positive even in the presence of a perturbation. On the other hand, a stock is robust if the stock return is not affected by the perturbation. Finally, a stock is fragile if the stock return decreases in the presence of the perturbation. These considerations apply both to stocks and cryptocurrencies.

### 3.3 Normalization for returns and perturbations of stocks and cryptocurrencies

Throughout this study two different systems were considered: the stock market and the cryptocurrency market. We used in each case different perturbation measures based on the available data. However, we consider the return (price differences) as a measure of satisfaction in both cases.

Since we are interested in measuring and comparing the antifragility in different systems, we normalized the stock return ***S(i)*** and the perturbation ***P(i)*** in such a way that both quantities are bounded in the intervals [-1, 1] and [0,1], respectively.

For the stock return ***S***, we use a new normalized return ***S'*** as

$$S' = \frac{S}{max(|S|)} \tag{6}$$

where **|*S*|** is the absolute value of ***S***, and ***max*(|*S*|)** is its maximum. With the definition given by Eq. (3), ***S'*** is a real number in the interval [-1,1].

For the perturbation ***P***, we use a new normalized perturbation ***P'*** as

$$P' = \frac{P - min(P)}{max(P) - min(P)}. \quad (7)$$

Clearly, ***P'*** is a real number in the interval [0,1].
These normalizations imply that the antifragility $A'=S'\cdot P'$, is a real number in the interval [-1,1].
In the following sections, we will assume that all the perturbations and returns have been already normalized in this way. We will also assume that the quantities under consideration such as the price $p_i$, the volume $v_i$ (the number of shares that are sold or traded), and the market capitalization $m_i$ (the market value as of a publicly traded company's outstanding shares) at time ***i***, have also been normalized to be in the interval [0,1].

### 3.4 Perturbation Measures

We considered perturbations coming from different sources for the stock market and for the cryptocurrencies market, according to the data we analyzed, all of them internal in accordance with the discussion in Section 2.2.2.
Essentially, we associate perturbations with fluctuations or volatility. So, we computed this measure by assessing how much the assets changed their parameters (price and volume in the case of stocks, and also market capitalization in the case of cryptocurrencies) in two consecutive observations.
In the case of stocks, as our data did not include values of market capitalization, we included other indicators, such as the volatility index, with symbol VIX, and the mean of the volatility of Nasdaq, Dow Jones and S&P 500 indexes.

More concretely, let $p(x)_i$, $v(x)_i$ and $m(x)_i$ be the price, volume, and market capitalization of an asset ***x*** at time ***i***, respectively. Then, in the case of stocks, we defined the different perturbations as:

- $P_{pr}(i) = mean_x P_{pr}(x, i)$, where $P_{pr}(x, i) = |p(x)_i - p(x)_{i-1}|$, (8)
- $P_{vl}(i) = mean_x P_{vl}(x, i)$, where $P_{vl}(x, i) = |v(x)_i - v(x)_{i-1}|$, (9)
- $P_{np}(i) = mean_x P_{np}(x, i)$, where $P_{np}(x, i) = |S(x, i)|$. (10)
- $P_{vix}(i)$ is the value of the volatility index VIX at time $i$.
- $P_{3m}(i)$ is the average of absolute changes of the Nasdaq, Dow Jones and S&P 500 indexes between time ***i-1*** and time ***i***. That is, if ***N(i), D(i), &(i),*** are the normalized values corresponding to such indices at time ***i***, then:

$$P_{3m}(i) = \frac{1}{3}\cdot[\,|N(i) - N(i-1)| + |D(i) - D(i-1)| + |\&(i) - \&(i-1)|] \quad (11)$$

Finally, the perturbation measure for stocks is defined as the mean value of these four perturbation sources:

$$P_{stk}(i) = \frac{1}{5} \left( P_{pr}(i) + P_{vl}(i) + P_{np}(i) + P_{vix}(i) + P_{3m}(i) \right). \quad (12)$$

In the case of cryptocurrencies we considered:

- $P_{pr}(i) = mean_x P_{pr}(x,i)$, where $P_{pr}(x,i) = |p(x)_i - p(x)_{i-1}|$, (13)
- $P_{vl}(i) = mean_x P_{vl}(x,i)$, where $P_{vl}(x,i) = |v(x)_i - v(x)_{i-1}|$, (14)
- $P_{mk}(i) = mean_x P_{mk}(x,i)$, where $P_{mk}(x,i) = |m(x)_i - m(x)_{i-1}|$, (15)
- $P_{np}(i) = mean_x P_{np}(x,i)$, where $P_{np}(x,i) = |S(x,i)|$. (16)

The values of $\boldsymbol{P_{pr}(x,i)}$ and $\boldsymbol{P_{np}(x,i)}$ are different in the sense that, for $\boldsymbol{P_{np}(x,i)}$, the prices were normalized before taking the mean over all the assets and, for $\boldsymbol{P_{pr}(x,i)}$, afterwards. So the perturbation contributed by an asset in the former measure is equal to another asset of different price if they changed the same percentage of their price. While in the latter, the perturbation of the system is governed by expensive cryptocurrencies.

Correspondingly, the definition of the perturbation measure is defined as the mean value of these four perturbation sources.

$$P_{crp}(i) = \frac{1}{4} \left( P_{pr}(i) + P_{vl}(i) + P_{mk}(i) + P_{np}(i) \right). \quad (17)$$

**3.5 Computation of antifragility**

We computed antifragility values for every asset in each system considering different time-scales: daily, weakly, monthly and yearly – the latter only plotted in the case of stocks. As we will see in the following sections, the behavior of our antifragility measure does not change drastically between time-scales. This observation is consistent with previous findings on the distribution of stock and cryptocurrencies returns [38,48-50]. We calculate the antifragility using Eqs. (1) and (2), considering the perturbation given by Eq. (12) (for stocks) and Eq. (17) (for cryptocurrencies).

**3.6 Performance measures**

We compared quantitatively the antifragility values with different "performance measures" listed next:

- **age**: age of the asset in days.
- **pct_dlt-pr**: maximum price minus minimum price divided by the mean price.
- **pct_dlt-mk**: same as before but using market capitalization instead of price.
- **pct_dlt-vl**: same as before but using volume.

- **pct_pr-f-i**: final price minus initial price divided by the mean price.
- **pct-mk-f-i**: same as before but using market capitalization instead of price.
- **pct-vl-f-i**: same as before but using volume.
- **pr_mea**: mean price.
- **pr_std**: price standard deviation.
- **mk_mea**: mean market capitalization.
- **vl_mea**: mean volume.

We recall that the variables which involve the market capitalization were only used in the case of cryptocurrencies, as we do not have this data for stocks. Instead, we used lists of the best stocks of the year (from 2010 to 2017) according to different stock market analysts, such as Forbes, Yahoo Finance, Stock Market Watch, among others. We will often refer to the stocks belonging to such lists as the 'top-performers', and to the variables listed above as the 'good-performance' measures.

## 4. Results and Discussion

As an example, in Fig. 1 we show, for two different cryptocurrencies (ethereum and bitcoin), a time series for perturbation, satisfaction (return) and antifragility. Notice that the time scales are different. We used normalized values as explained above.

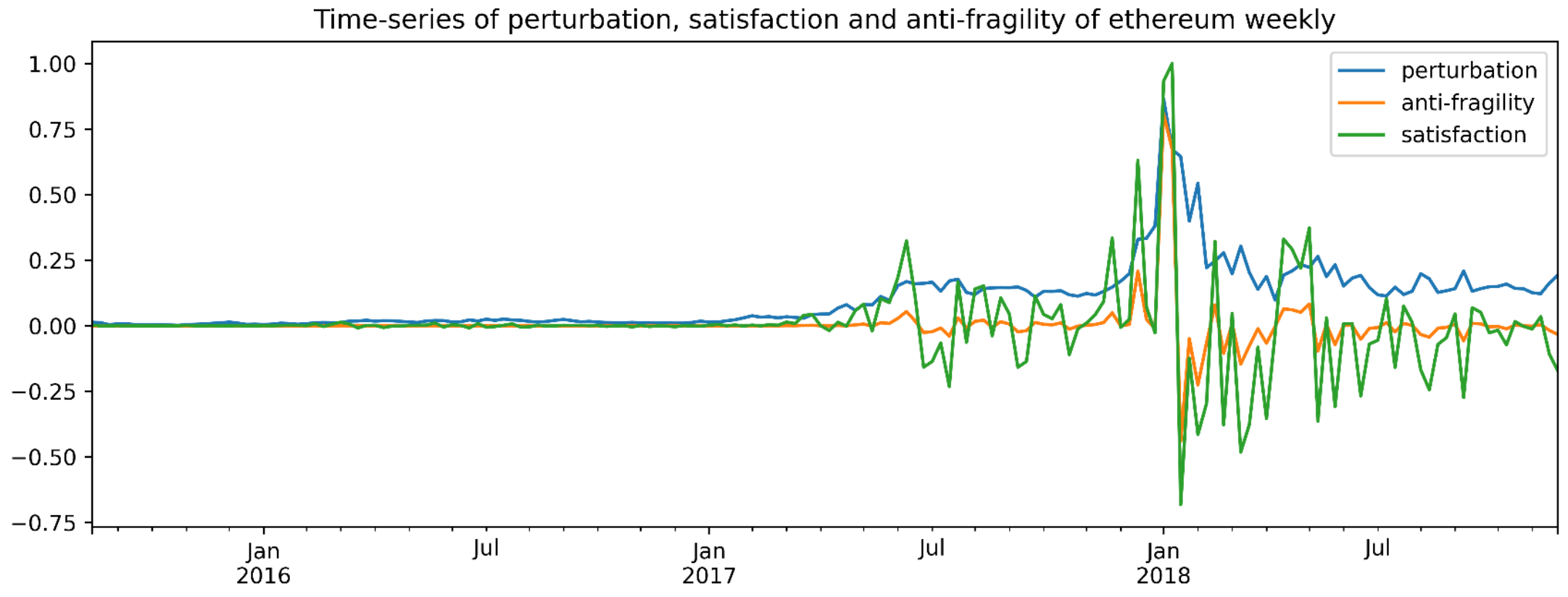

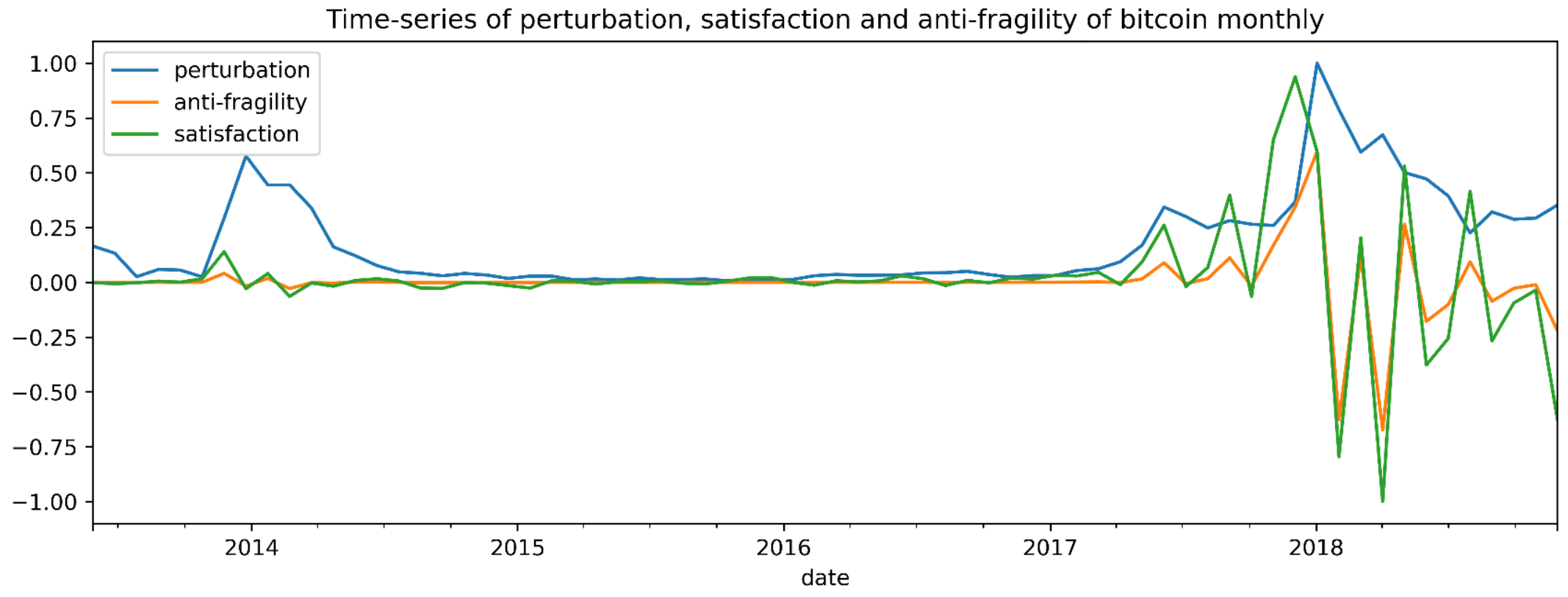

Figure 1: Time-series of the values of perturbation, satisfaction and anti-fragility of ethereum and bitcoin cryptocurrencies in weekly and monthly intervals, respectively. Notice that the time intervals are different.We calculated the satisfaction (price return) using Eq. (1) for the discrete time ***i=1,2,...,N***; and the antifragility using Eq. (2). For the perturbation in Eq. (2) we use the average quantity for cryptocurrencies given by Eq. (17). We plot this perturbation as a third curve.

### 4.1 Antifragility in the stock and cryptocurrency markets

In what follows, we will present a set of figures where we analyze the data of assets for the stock and cryptocurrency markets. We will show the values of antifragility versus the different performance measures listed above in Section 3.6.

As a first example, we plot in Fig. 2 the antifragility ***A***, that we indicated as "aft", against one of the performance measures. In this case, we used the mean price (pr_mean) of several cryptocurrencies. The mean price is evaluated considering the full range of the discrete time in the data set. We consider weekly prices here. We are plotting a green dot for each type of cryptocurrency and we have around 1800 different cryptocurrencies in the data set. We only indicate in the figure the name of some of the cryptocurrencies for clarity. It is worth remembering here the meaning of antifragility ***A***: we consider that an asset is robust if ***A*** is close to 0, fragile if it is negative, and antifragile if it is positive. Notice that the dots are centered around zero with a considerable dispersion.

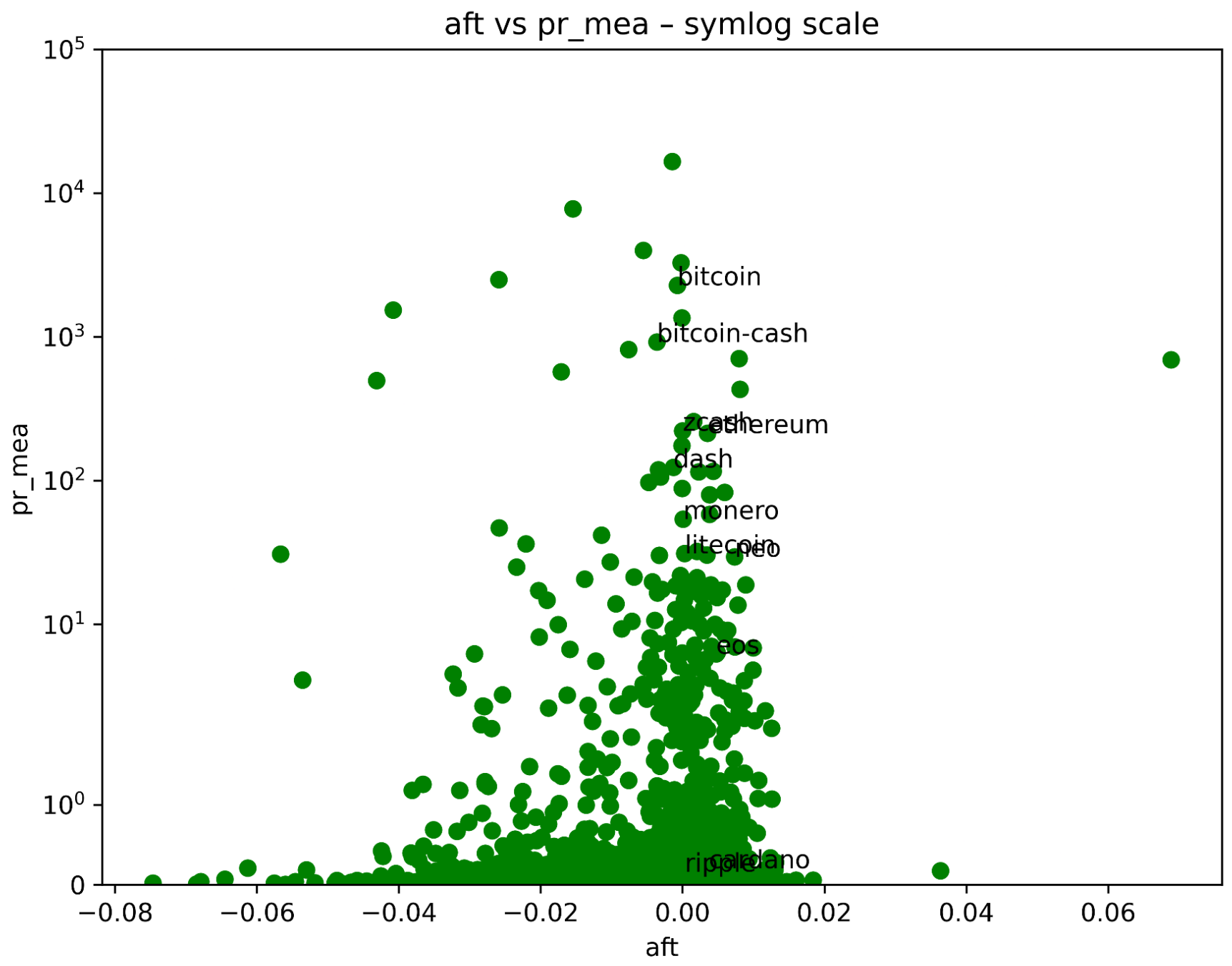

Figure 2: A typical example of the different types of studies that we analyze. We compare the antifragility values (in the $x$-axis) of every cryptocurrency in a weekly scale (according to the green color), with its mean price, indicated by pr_mea, (in the $y$-axis). Note that in this case the symlog scale is just a log scale because there are no negative values in the $y$-axis.

We now show a series of similar figures of antifragility *vs* different performance measures. We will consider different time scales for the antifragility, distinguished with a corresponding color: daily in blue, weekly in green, monthly in red and semiannual (only included in the case of stocks) in yellow. In Figs. 3 and 5, the horizontal axis represents antifragility and the value of $\boldsymbol{A}$ is centered around zero. In these figures, each dot represents a different asset in the stock market, and we indicate inside each figure just a few of their names for clarity. The number of stocks analyzed is about 7000.

We will start with Fig. 3(a,b,c,d) that contains four plots: each one representing a different time scale. In Fig. 3(a), we are plotting the age (measured in days) of the asset in question. In Fig. 3(b), we show the maximum volume minus the minimum volume divided by the mean volume (indicated by "pct_dlt-vol"). In Fig. 3(c), we plot the mean price, and in Fig. 3(d), we show the mean volume.
About this figure we can say the following:

1. The qualitative behavior of the antifragility $\boldsymbol{A}$ is the same for different times scales: daily (blue), weekly (green), monthly (red) and semiannual (yellow).
2. There are no linear correlations between any of the 'good-performance' measures and the antifragility measures.
3. The higher values of the 'good-performance' measures concentrate around the robust ($\boldsymbol{A=0}$) axis. We used different scales in order to better illustrate this feature.
4. The behavior pointed out in (3) is clearer and more consistent for different time scales among the 'good-performance' measures in all of these figures.

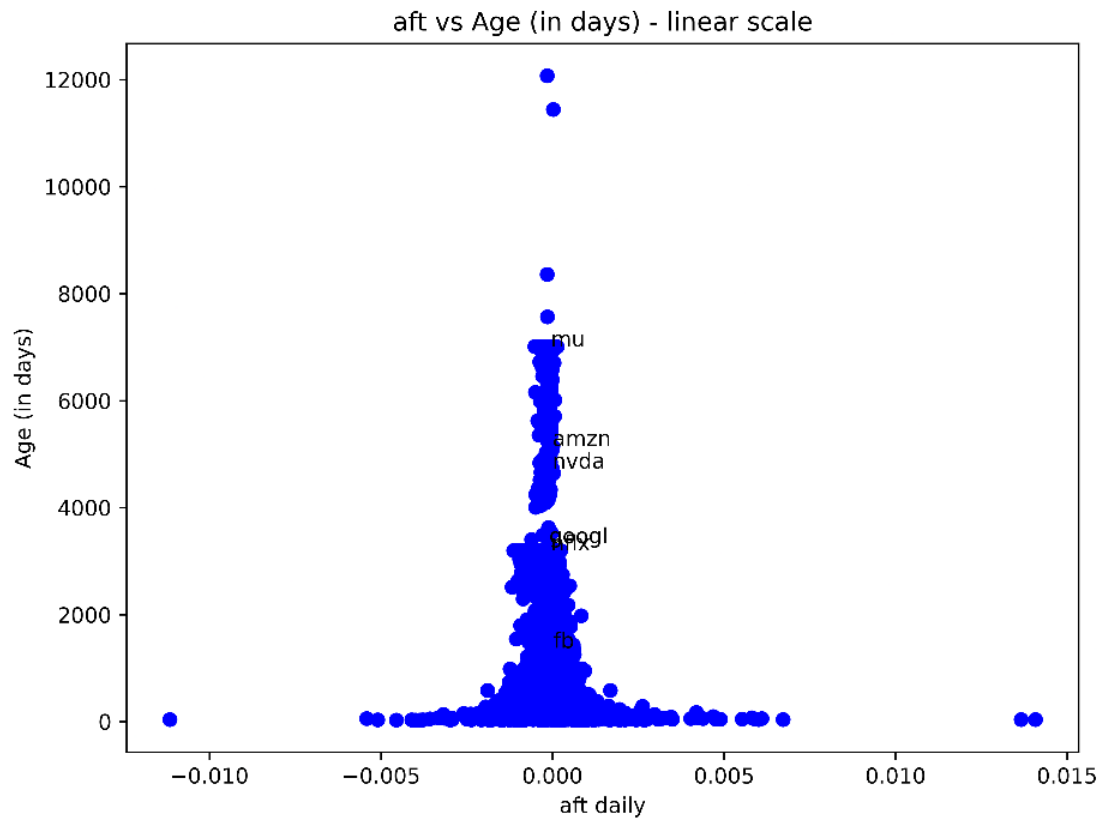
aft vs Age (in days) - linear scale
Age (in days)
aft daily
mu
amzn
nvda
googl
fb

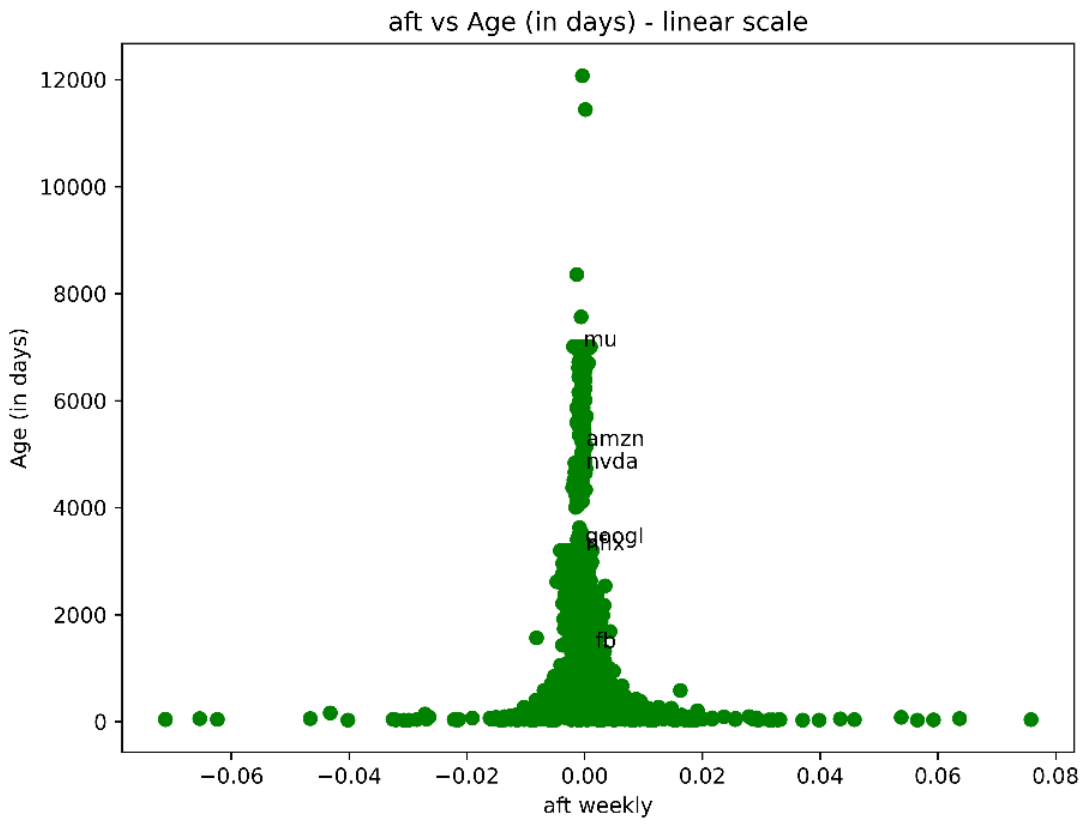
aft vs Age (in days) - linear scale
Age (in days)
aft weekly
mu
amzn
nvda
fb

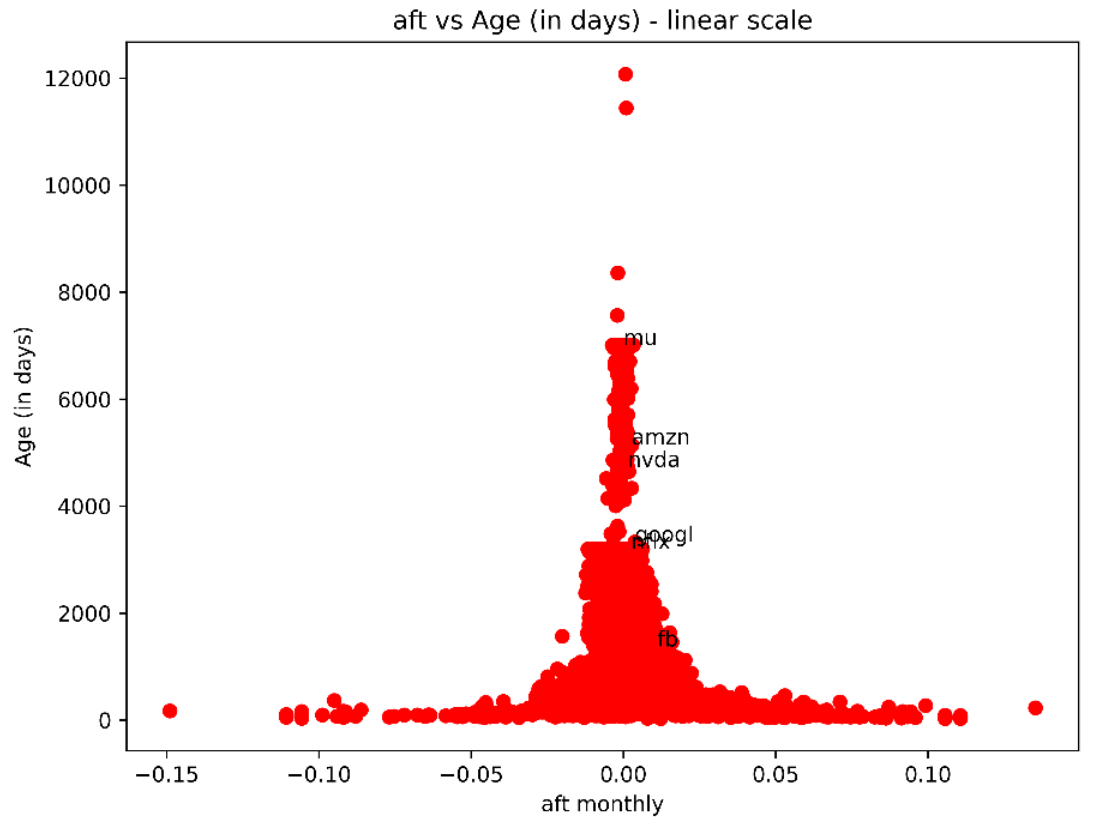
aft vs Age (in days) - linear scale
Age (in days)
aft monthly
mu
amzn
nvda
googl
fb

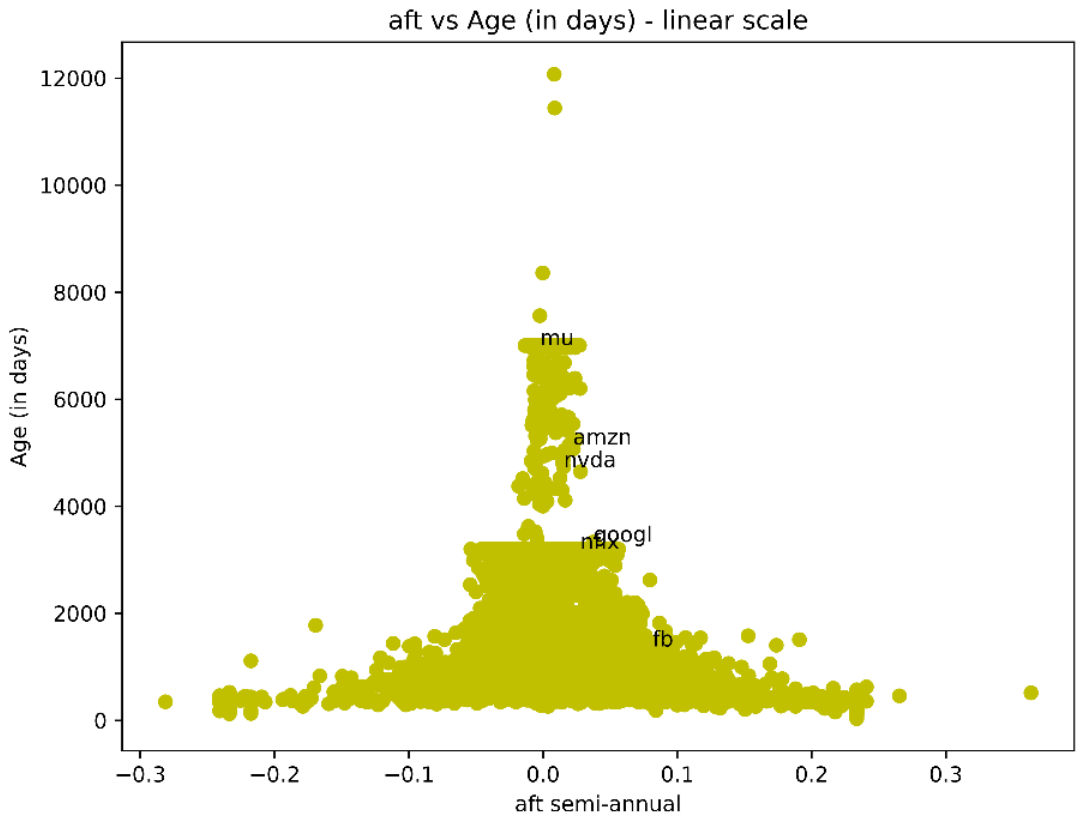
aft vs Age (in days) - linear scale
Age (in days)
aft semi-annual
mu
amzn
nvda
googl
fb

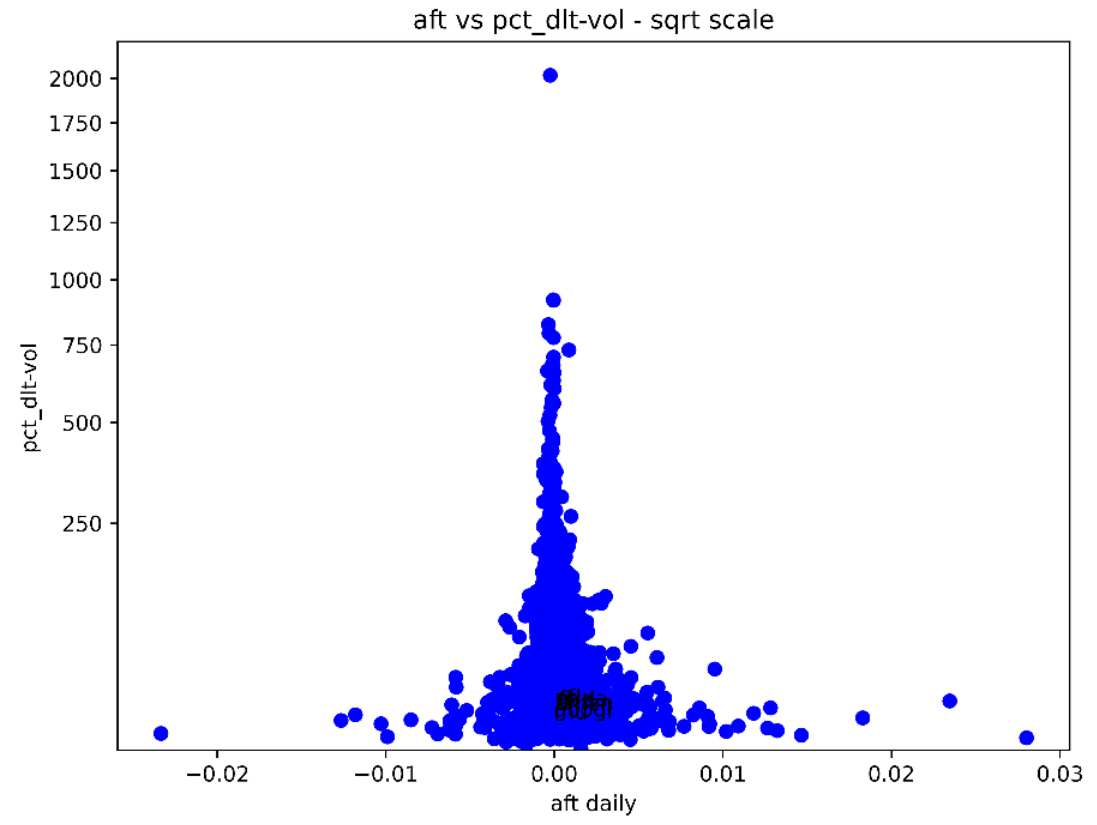
aft vs pct_dlt-vol - sqrt scale
pct_dlt-vol
aft daily

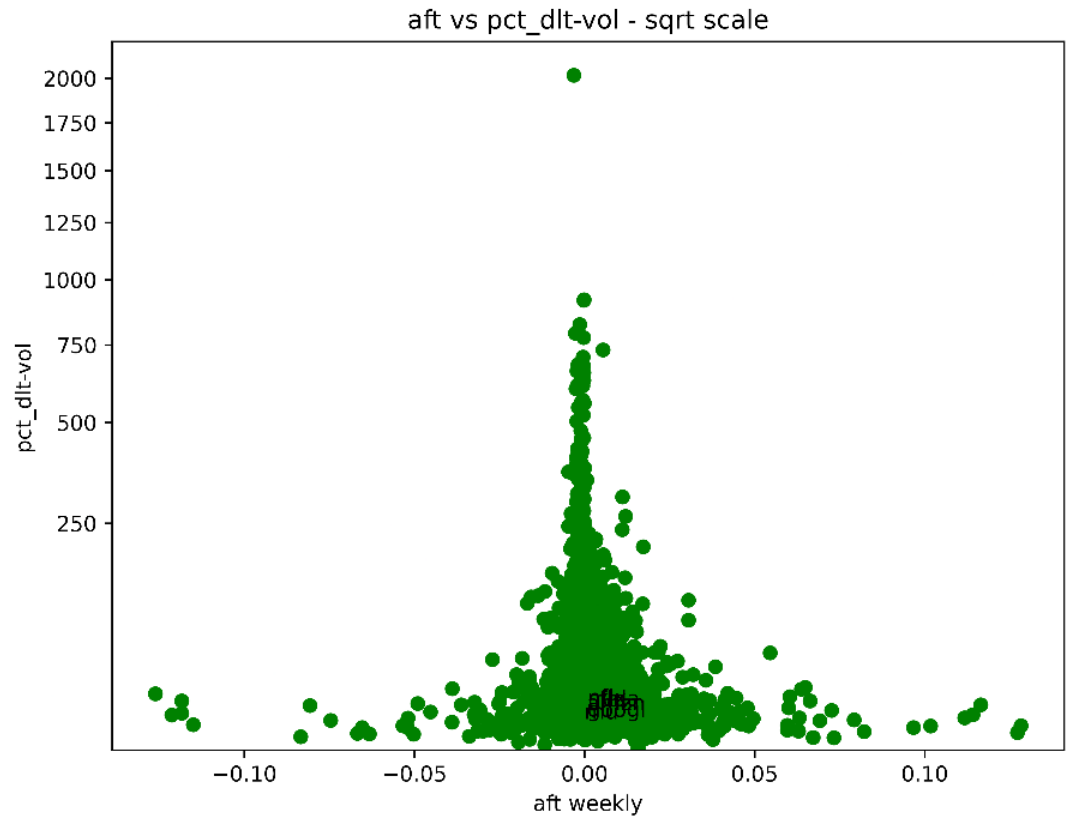
aft vs pct_dlt-vol - sqrt scale
pct_dlt-vol
aft weekly

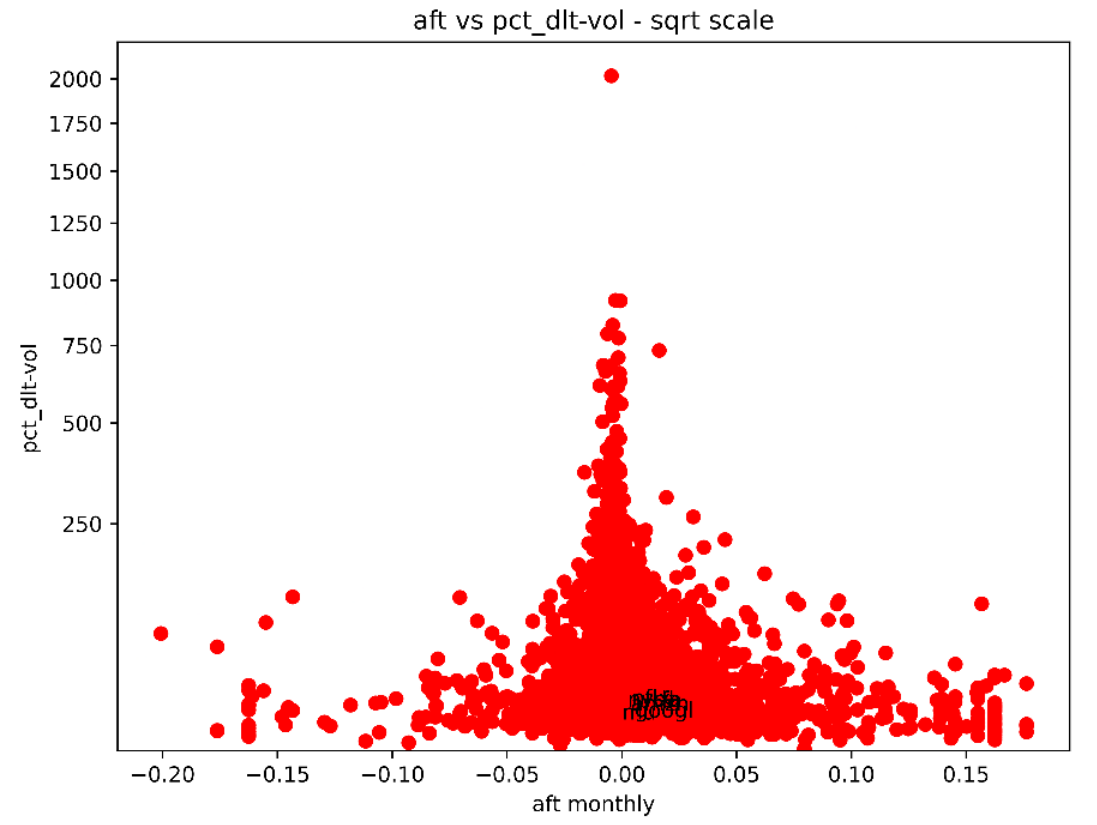
aft vs pct_dlt-vol - sqrt scale
pct_dlt-vol
aft monthly

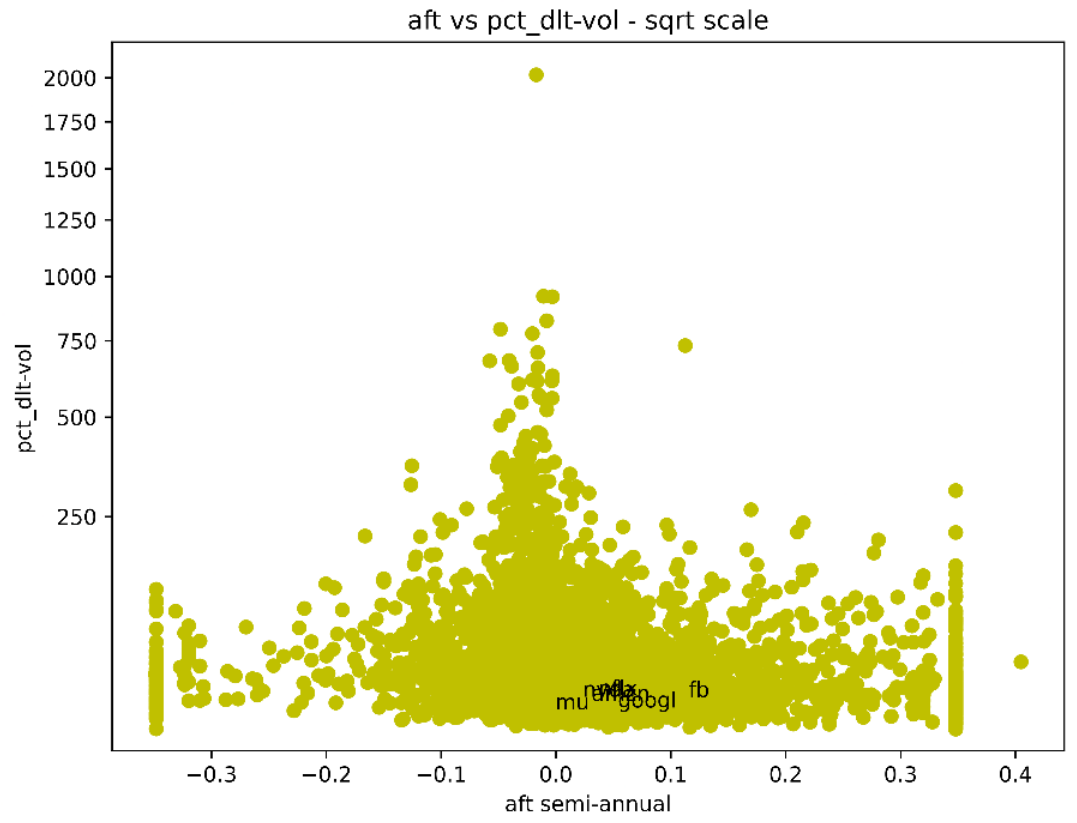
aft vs pct_dlt-vol - sqrt scale
pct_dlt-vol
aft semi-annual
mu
googl
fb

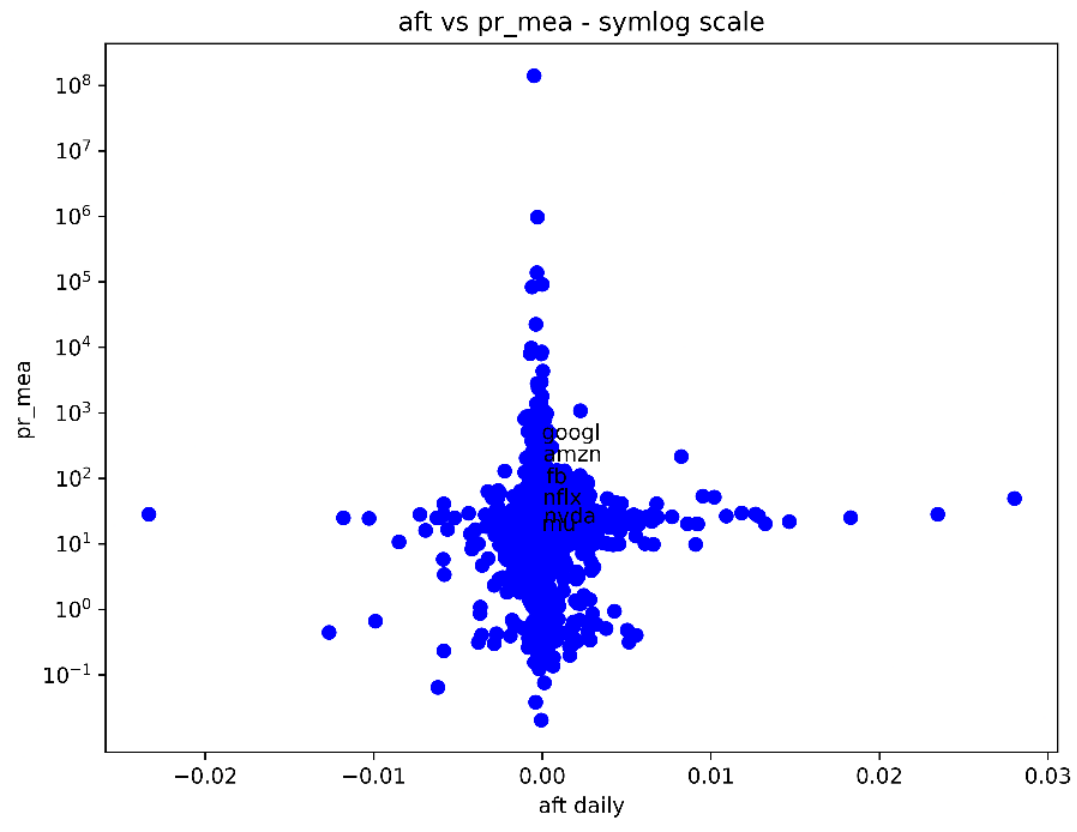
aft vs pr_mea - symlog scale
pr_mea
aft daily
googl
amzn
fb
nflx
nvda
mu

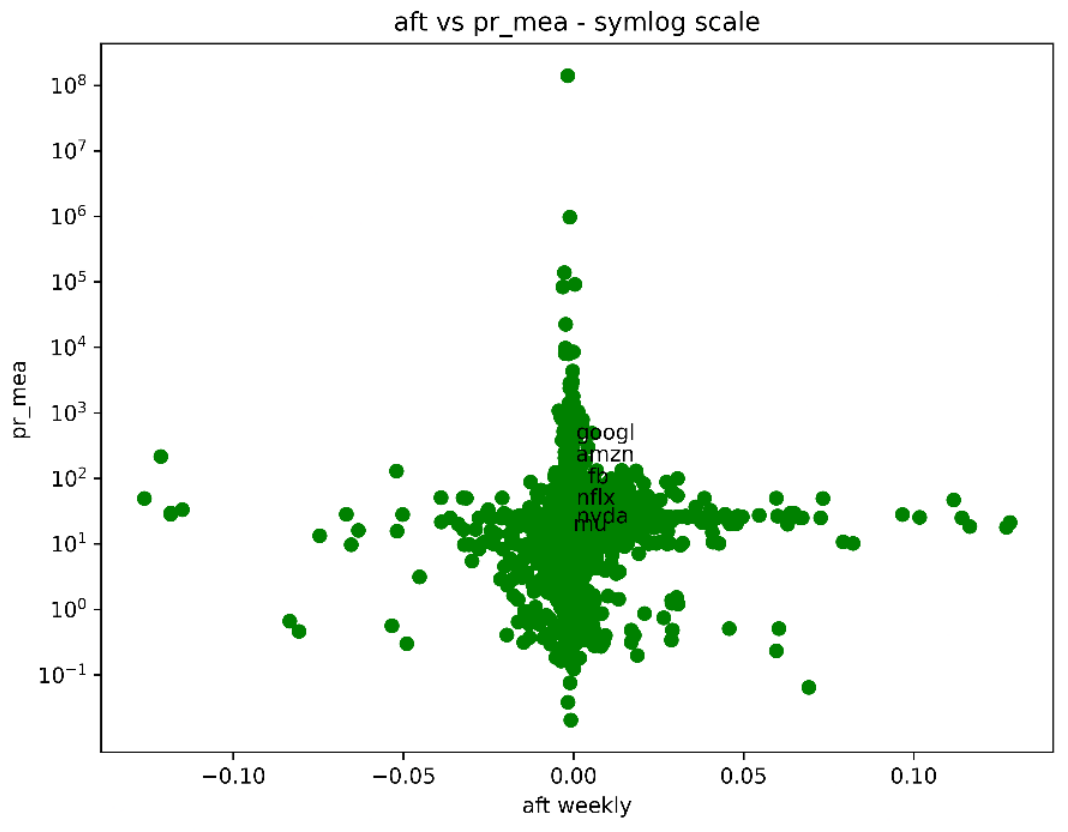
aft vs pr_mea - symlog scale
pr_mea
aft weekly
googl
amzn
fb
nflx
nvda
mu

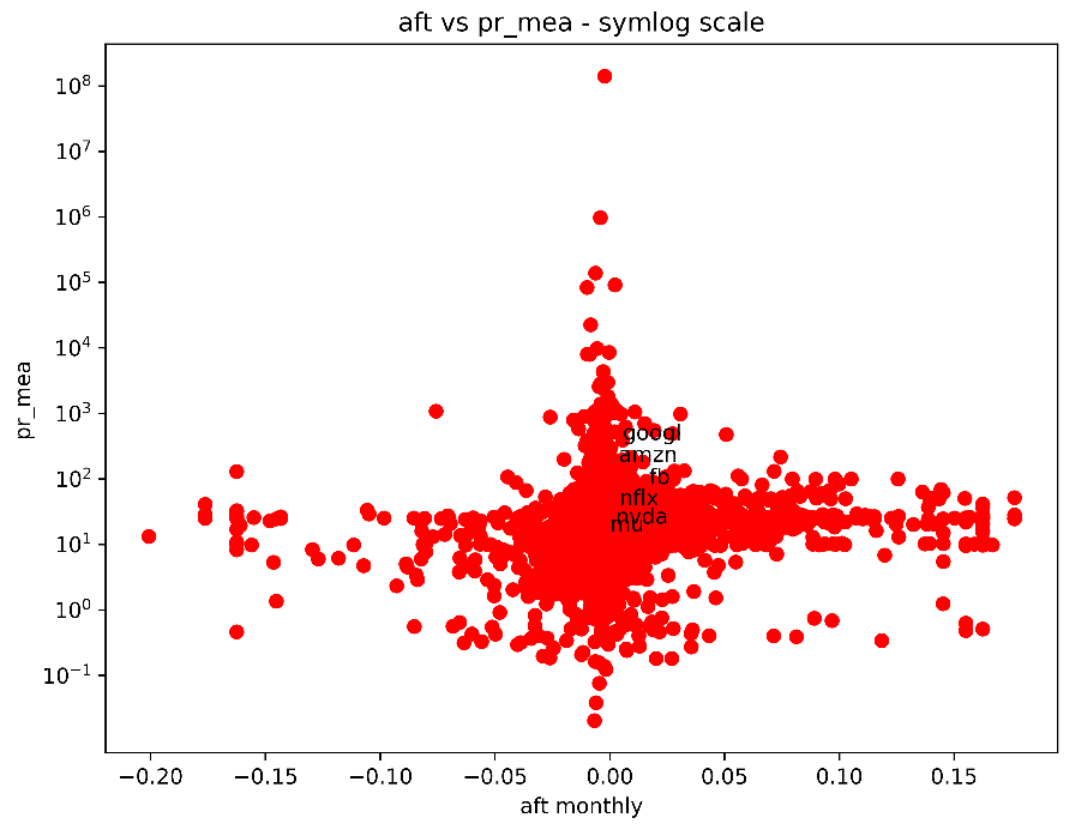
aft vs pr_mea - symlog scale
pr_mea
aft monthly
googl
amzn
fb
nflx
nvda
mu

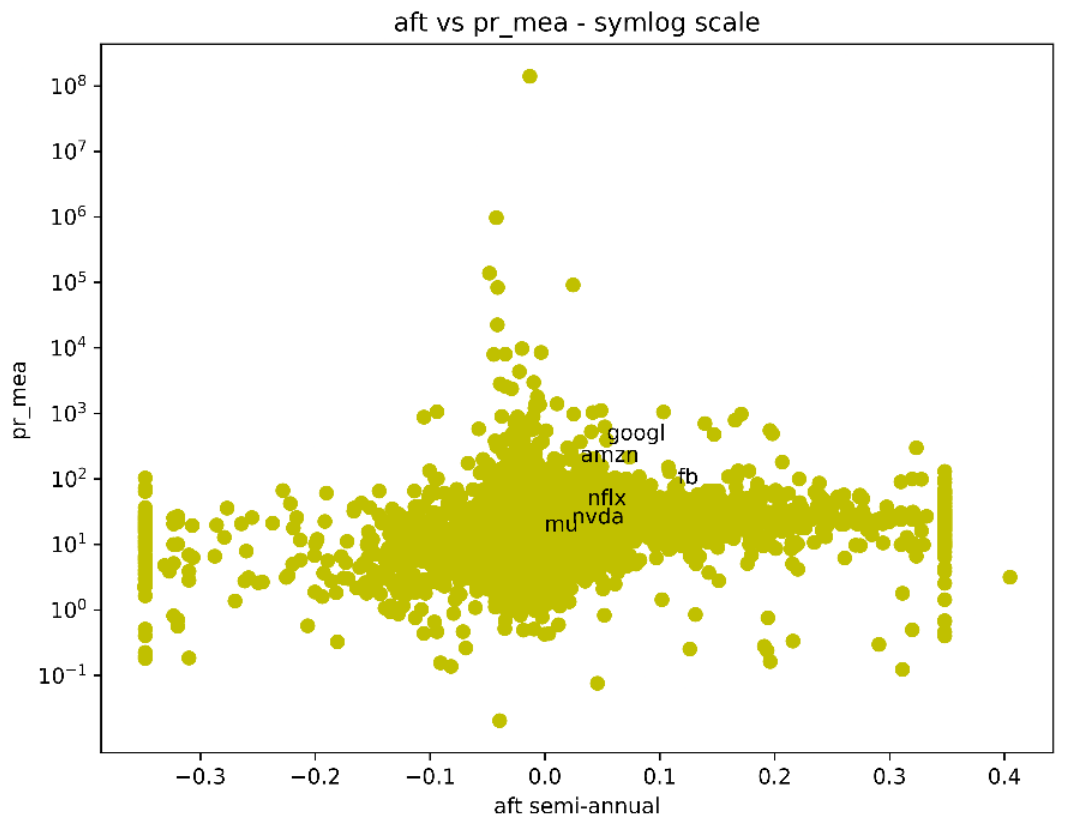
aft vs pr_mea - symlog scale
pr_mea
aft semi-annual
googl
amzn
fb
nflx
nvda
mu

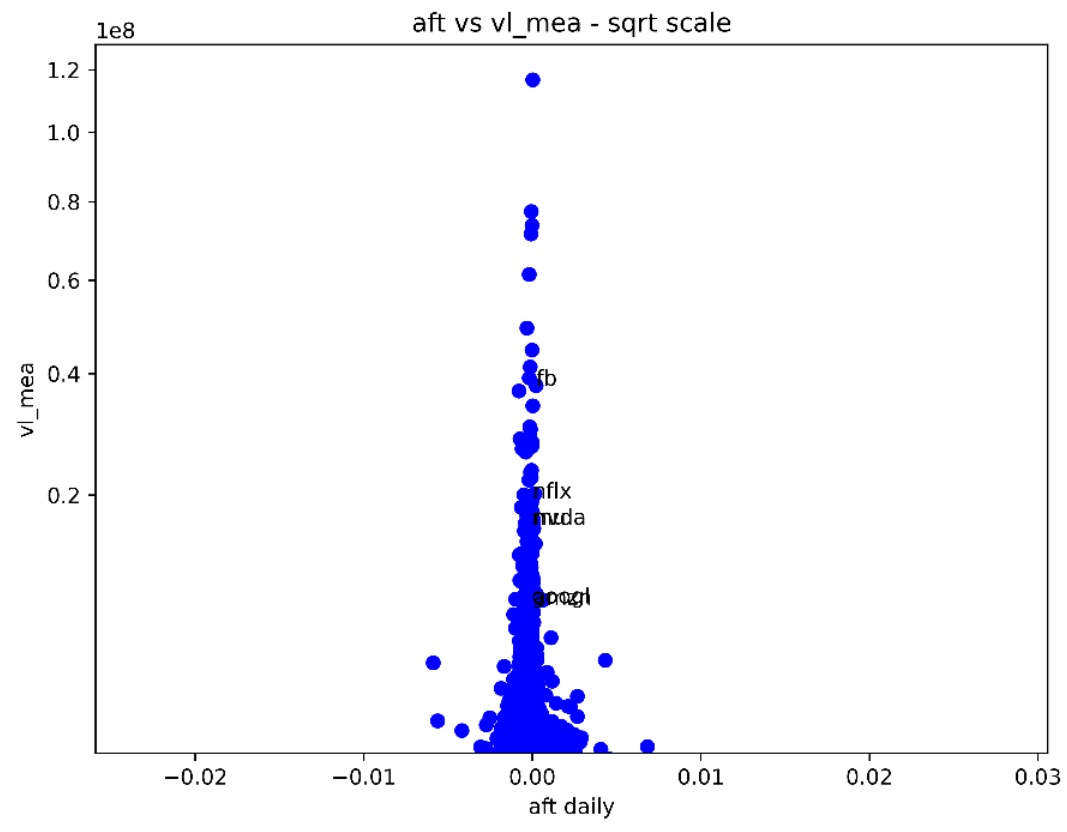
aft vs vl_mea - sqrt scale
1e8
vl_mea
aft daily
fb
nflx

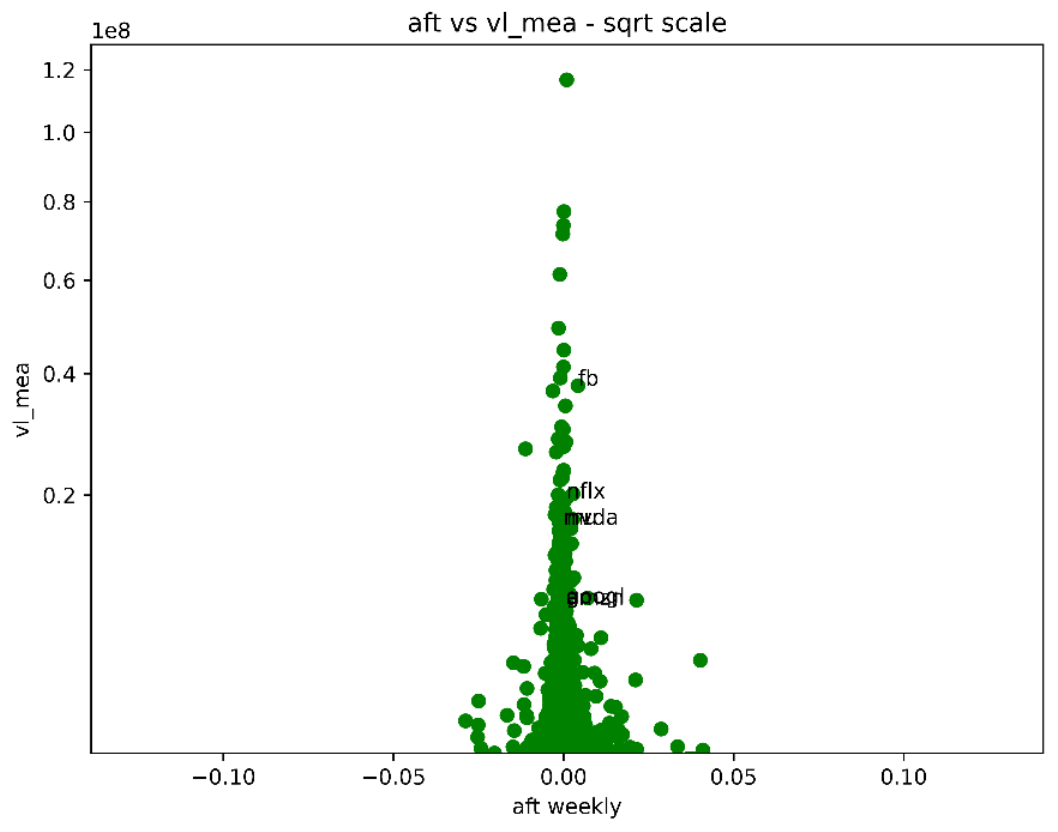
aft vs vl_mea - sqrt scale
1e8
vl_mea
aft weekly
fb
nflx

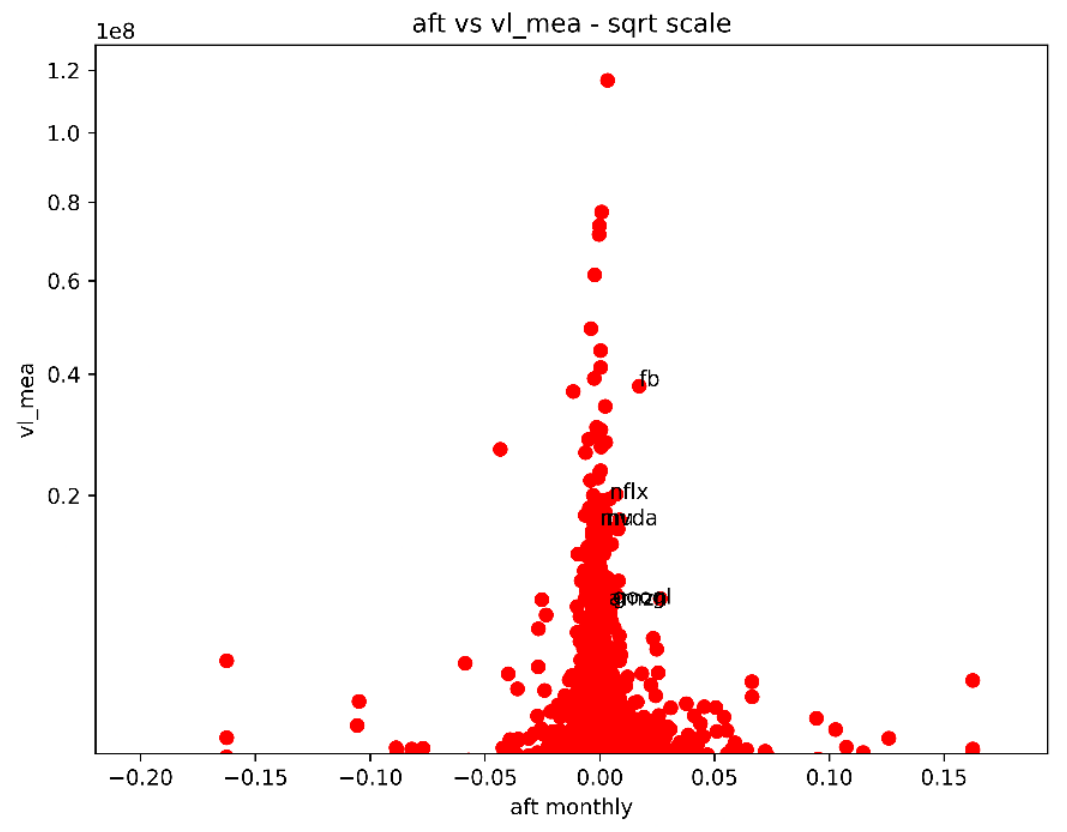
aft vs vl_mea - sqrt scale
1e8
vl_mea
aft monthly
fb
nflx

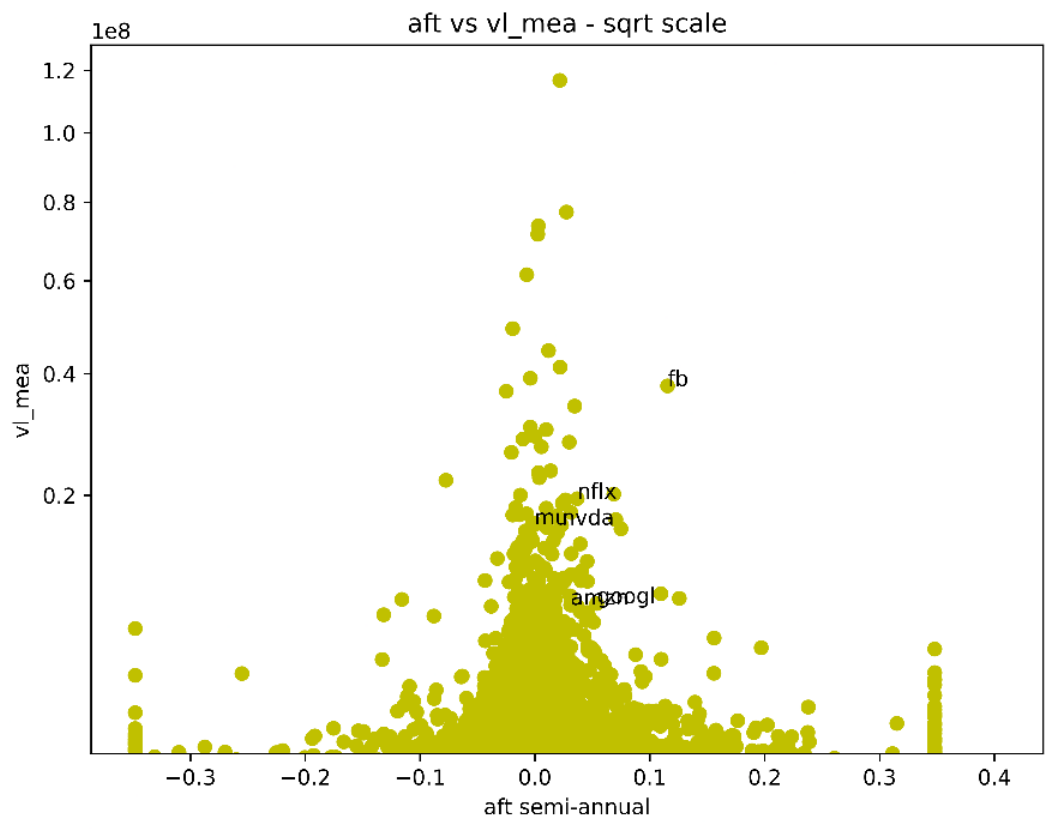
aft vs vl_mea - sqrt scale
1e8
vl_mea
aft semi-annual
fb
nflx
munvda

Figure 3: Comparison of antifragility values and 'good-performance' measures in stocks. There are four images in the figures (a,b,c,d), corresponding to a 'good-performance' measure: **age, pct_dlt-vol, pr_mea, vl_mea**. Each subplot corresponds to a timescale (daily in blue, weekly in green, monthly in red, semiannual in yellow). The plots compare the antifragility ***A*** measure (in the *x*-axis) to the corresponding 'good-performance'. We observe how higher values of the 'good-performance' measure concentrate in the robust axis $\boldsymbol{A} = \mathbf{0}$. See the text for details of this figure.

We now plot, in Fig. 4, the probability distributions of the antifragility for different time scales, indicated by different colors as shown in the inset. Notice that these distributions are close to normal distributions.

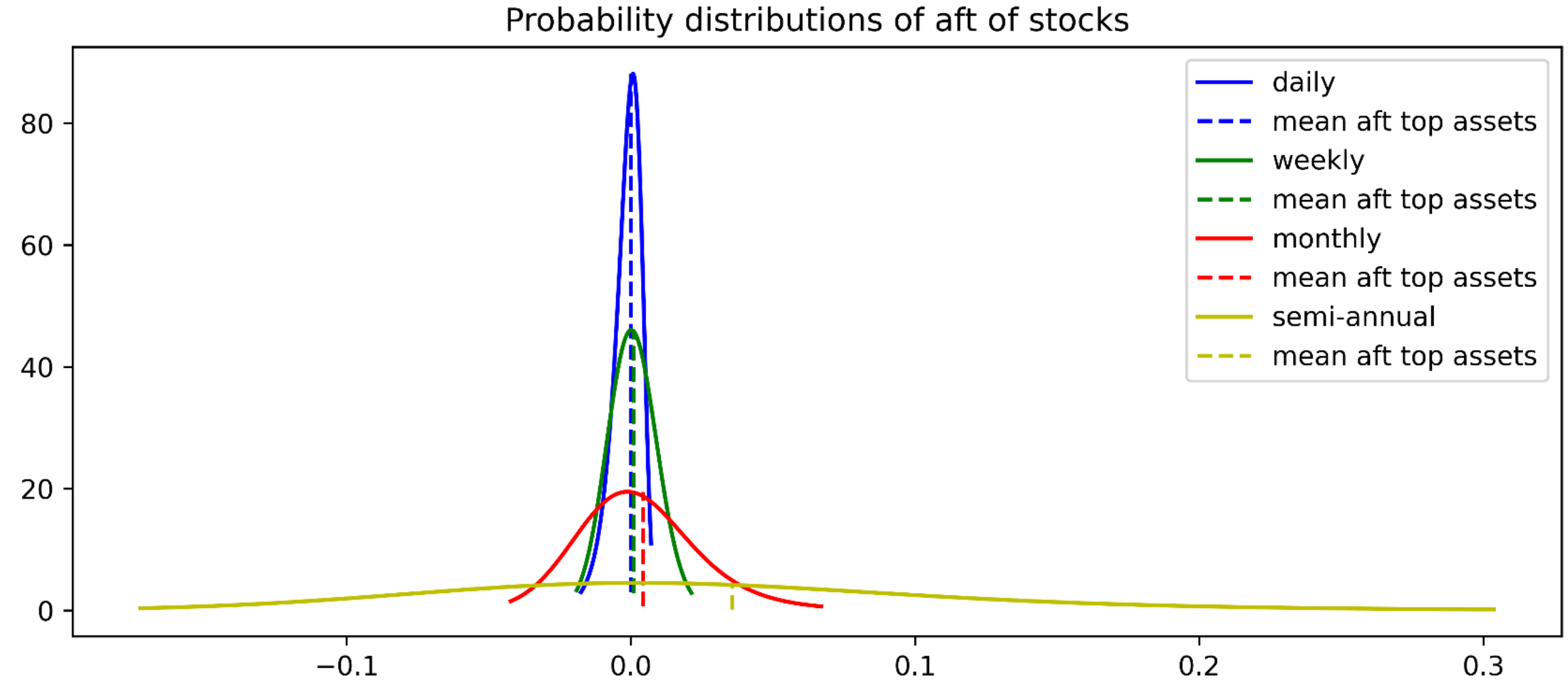

Figure 4: Probability distribution of the antifragility ***A*** for stocks for different time-scales. The dashed line represents the mean antifragility of the set of 'top-performers'. We observe that the mean ***A*** of 'top-performers' is only greater than the mean ***A*** in the daily timescale.

Now we turn our attention to the case of the cryptocurrency market.
In Fig. 5, we carried on essentially the same analysis as in the case of stocks. Among the following plots, colors represent a time scale: daily in blue, weekly in green, monthly in red. In Fig. 5(a), we are plotting the age that corresponds to the age in days of the asset in question. In Fig. 5(b), we show the maximum volume minus the minimum volume divided by the mean volume (indicated by "pct_dlt-vol"). In Fig. 5(c), we plot the mean price. Finally, in Fig. 5(d), we plot the same as in (b) but with price instead of volume (indicated by "pct_dlt-pr").

We observe a similar behavior as in the case of stocks. The previous observations (1-4) also hold in this case. And it can be observed a greater concentration of assets on the 'fragile' side.

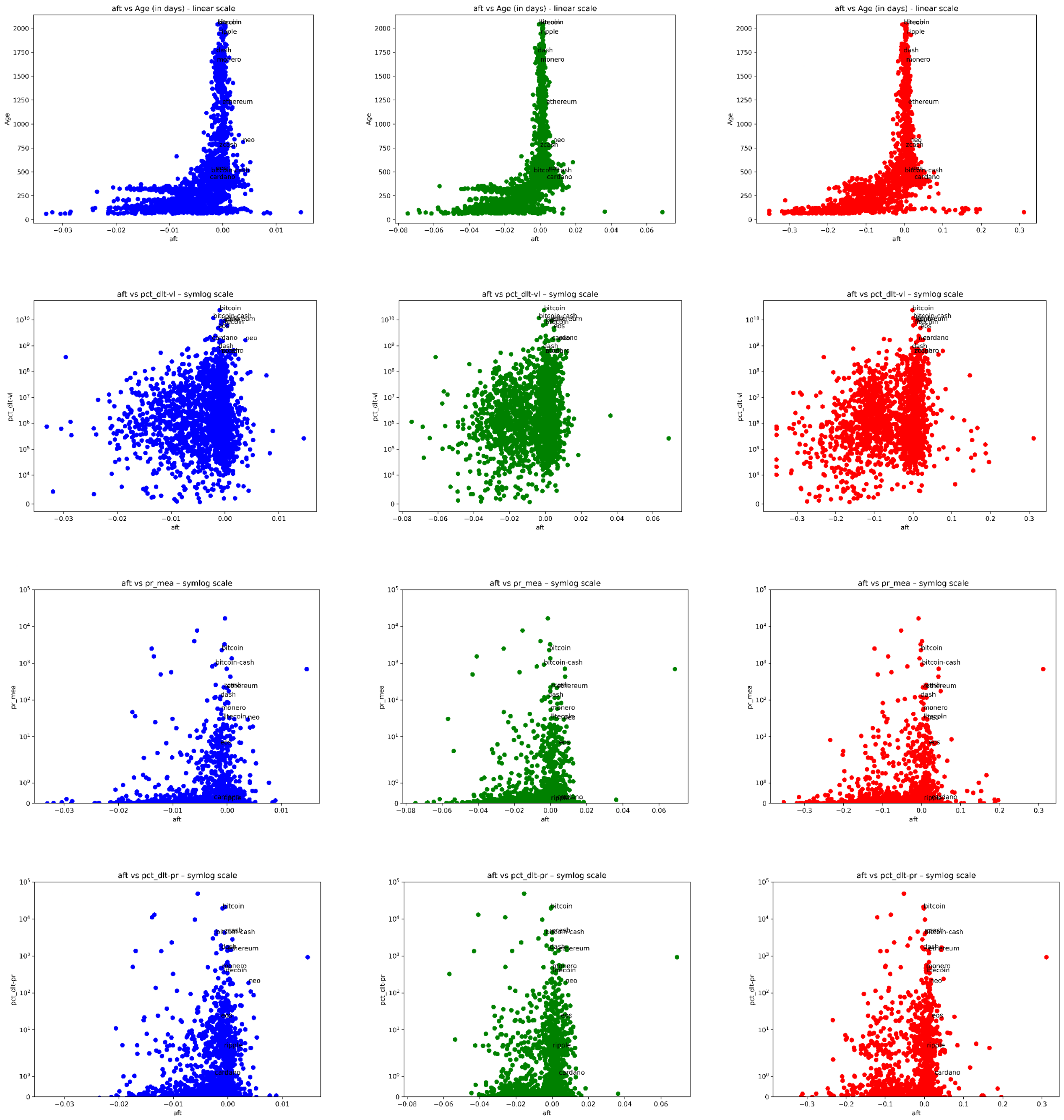

Figure 5: Comparison of anti-fragility values and 'good-performance' measures in cryptocurrencies. There are three frames in the image, each of them corresponds to one 'good-performance' measure, from up to bottom: **age, pct_dlt-vl, pr_mea, pct_dlt-pr**. See text. Each subplot of these frames corresponds to a timescale. Each plot compares the corresponding $\boldsymbol{A}$ measure (x-axis) to the 'good-performance' measure under consideration. We observe how higher values of the 'good-performance' measure concentrate in the robust axis $\boldsymbol{A}$**=0**.

Finally, the probability distributions of the antifragility measures, shown in Fig. 6, are close to skew-normal distributions, with a left asymmetry towards lower values of antifragility. The probability distributions of the antifragility values of the 'top-performers' seem to slightly improve from the rest. See Fig. 6.

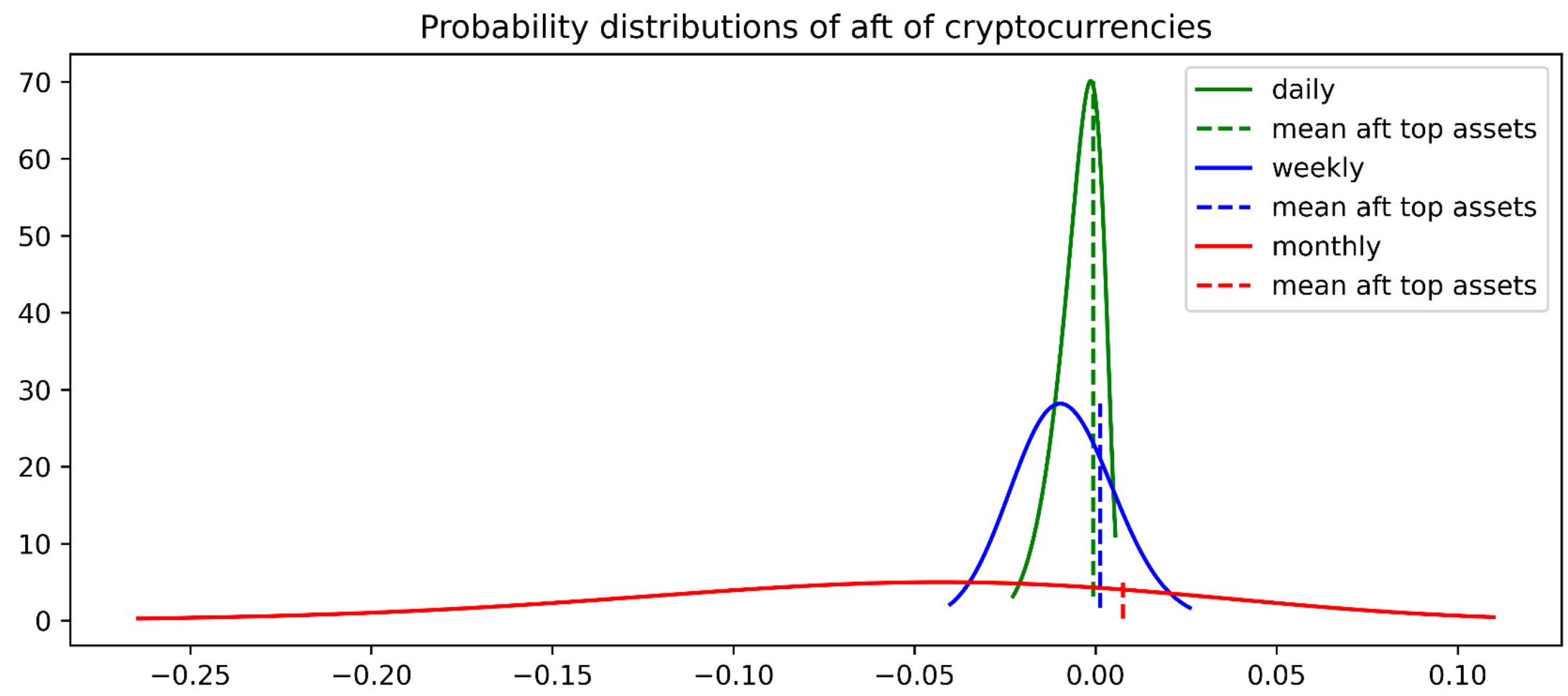

Figure 6: Probability distribution of the antifragility $\mathcal{A}$ of cryptocurrencies in different time-scales. The dashed line represents the mean antifragility of the set of 'top-performers'. We observe that the mean $\mathcal{A}$ of 'top-performers' is greater than the mean $\mathcal{A}$ in every timescale.

## 5. Discussion and Conclusions

Antifragility occurs when systems benefit from perturbations. This is opposite to fragility, where noise or changes damage a system. The absence of fragility, when perturbations do not affect a system, is robustness. In the introduction of this paper, we briefly mentioned some systems that benefit in the presence of noise. We mentioned some phenomena that have been thoroughly studied recently: stochastic resonance, noise-induced transport and optimization by simulated annealing, among others. All these phenomena share the property of gaining in the presence of a stochastic signal or perturbation, thermal noise or being exposed to other fluctuations.

As mentioned in the introduction, other important examples of complex dynamical systems that might have antifragility, are economic and financial systems. Clearly, both are complex systems that are open, nonlinear, and have many agents that operate collectively, giving rise to emergent properties like the ones we observe in financial markets and other economic indicators.

An important question that we may ask is: can we identify the phenomenon of antifragility in economics and finance? As far as we know, this question has not been answered in a definite way and, worse than that, the question usually is not even asked or posed in these terms. Therefore, the main goal of this paper is: firstly, to pose this question explicitly and, secondly, to try to answer it by proposing a novel measure of antifragility for stocks and cryptocurrencies markets.

Our proposal is to define explicitly the antifragility in mathematical terms as a multiplicative stochastic process. Our definition is simple: antifragility is the product of returns and perturbations. Based on this definition, we analyze the returns to be used: price returns of stocks and cryptocurrencies in the market. As perturbations, we used different options that come from the market itself, like fluctuations of price returns, volume, market capitalization and global average indexes. One can think of these perturbations as endogenous shocks. We have not studied here the case of exogenous shocks or aggregate shocks, but it could be interesting to analyze these cases in a further analysis for the future. Finally, with these returns and perturbations, we calculated the antifragility measure. Both the returns and the perturbations were obtained from data sets that are publicly available, as indicated in the text. Once we obtain the antifragility, we explore its dependence against different performance measures: average quantities related with price, volume, and market capitalization.
It is worth mentioning that, by defining antifragility in the way we did, there are first-order effects that can be seen in a set of assets whose antifragility and satisfaction (price returns) time series are highly correlated (either positively or negatively). This happens for different reasons, either they are very new assets and so there are few observations, or they had almost no change in some of their parameters such as a constant price or volume.

As a matter of conclusion of this study, we can point out the following:
The results obtained with our definition of antifragility indicate that both the stock and the cryptocurrencies markets do not show an antifragile character in a clear way and seem to be favored by robustness instead. There were no correlations between our antifragility measures and the different performance measures that we analyze, while the best performance assets seem to fall within the category of robustness. In virtually every analysis we carried out, the highest and lowest antifragility values were obtained for the newest assets and for those with the lowest transaction volume. For example, the average age of the top ~300 antifragile stocks is 302 days, the oldest of them is less than 3 years old, and over 80% are younger than a year. Moreover their mean volume of transactions is around 5 times smaller than the mean volume of transactions overall. It can be observed that, in both scenarios, the assets tend to evolve towards robustness. In Fig 7 we show this feature by plotting the cumulative mean of the antifragility measure over time. In Fig. 7(a), it is shown this evolution for a selected set of 'top-performing' stocks in a monthly scale for about twelve years; Fig. 7(b) shows the same time scale and span but with a random set of assets; and Fig. 7(c) shows a daily evolution of a selected set of cryptocurrencies for one year.

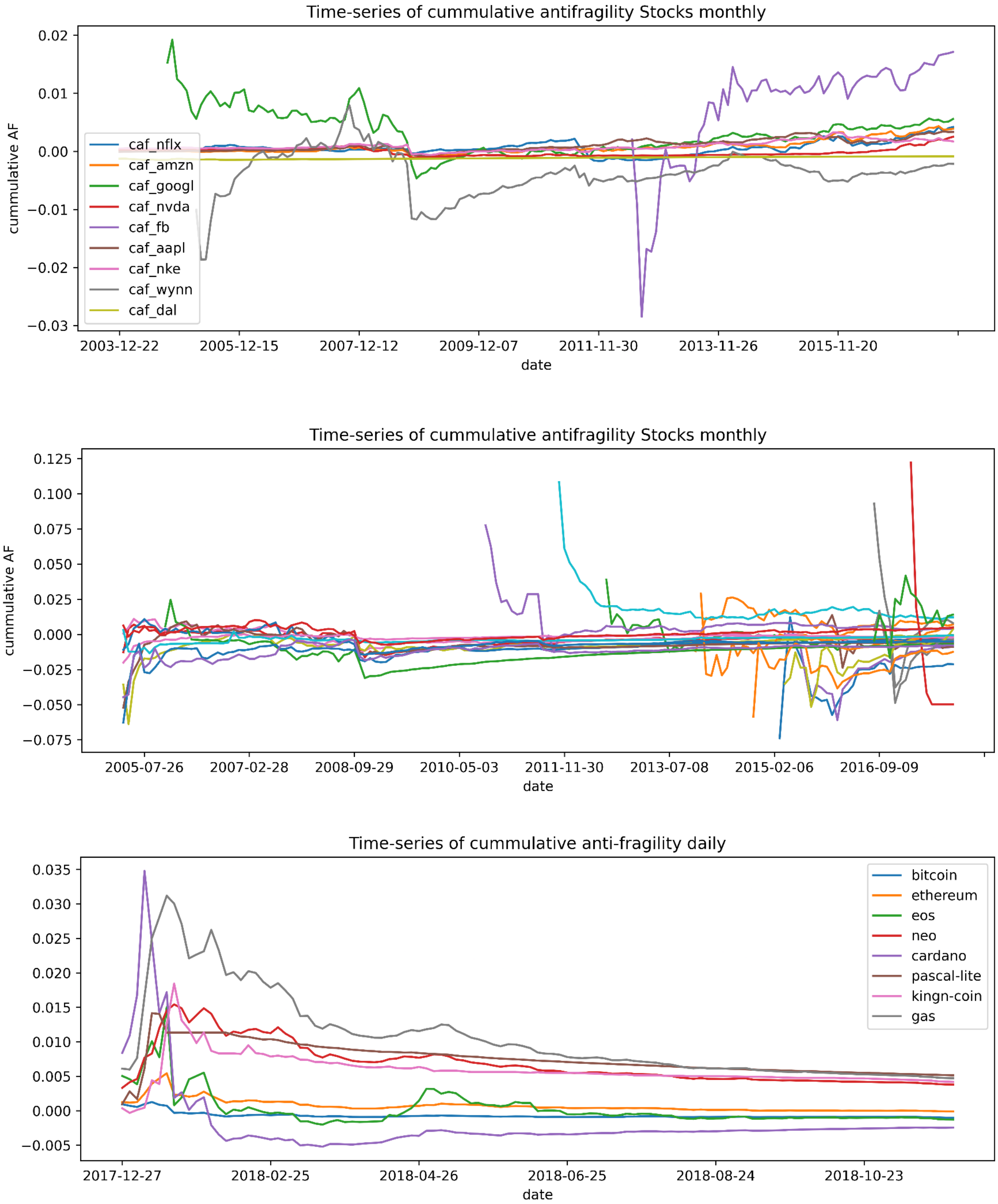

Figure 7: Cumulative mean of antifragility of few assets. The $x$-axes shows the date and the $y$-axis the cumulative antifragility measure. Figure (a) shows the cumulative monthly antifragility (expressed as 'caf' in the image) of selected 'top-performing' stocks , (b) shows the cumulative monthly antifragility of 25 random stocks, and (c) shows a daily evolution of a selected set of cryptocurrencies for one year.

Finally, it is important to stress that these results depend obviously on the definition of antifragility that we used. It would be interesting to address other possible definitions that capture the essence of the phenomenon. Meanwhile, we sincerely hope that this work stimulates the interest of researchers to study antifragility as an important feature in financial systems and in complex dynamical systems in general.

## 6. Data Availability

We analyzed a data set with historical daily prices and volumes of more than 7,000 stocks from the US Stock Market [51]. We truncated the data starting from 1990. This data set considered only active stocks by the end of 2017 and there is no data from stocks that disappeared from the market before that date.
In the case of cryptocurrencies, the data [52] consisted of historical daily data (open price, market capitalization and volume) of around 1800 assets, the oldest being from 2013 and up to November 2018. Instructions for accessing the processed data sets can be found in this purl: https://purl.prod.archive.org/purl/AFSCM_data/processed-data.

## 7. Conflicts of Interest

The authors declare that there is no conflict of interest regarding the publication of this paper.

### 7.1 Funding Statement

This project was supported by an award from the Fundación Marcos Moshinsky A. C. and UNAM's PAPIIT IN107919 and IV100120 grants.

## 8. Author contributions

CG: Conceptualization; Funding acquisition; Project administration; Supervision; Methodology; Validation; Writing - review & editing.
DA: Data curation; Formal analysis; Investigation; Methodology; Resources; Software; Visualization; Writing - original draft.
JLM: Formal analysis; Investigation; Validation; Writing - original draft; Writing - review & editing.

## 9. Acknowledgments

We are grateful to Ewan Colman, Hyobin Kim and Dante Pérez for interesting discussions and suggestions; and to the C3 community, specially to José Luis Gordillo for providing technical support with the Center's Cluster usage.